\documentclass[10pt]{article}
\usepackage{fullpage}
\usepackage{amsmath}
\usepackage{amssymb}
\usepackage{amsfonts}
\usepackage[dvips]{epsfig}
\usepackage{color}
\usepackage{cite}
\usepackage{upgreek}

\def\##1{{\underline{#1}}}
\def\=#1{\underline{\underline{#1}}}

\def\+#1{\underline{\bf #1}}
\def\*#1{\underline{\underline{\bf #1}}}

\def\r#1{(\ref{#1})}
\def\l#1{\label{#1}}
\def\c#1{\cite{#1}}

\def\le{\left(}
\def\ri{\right)}
\def\les{\left[}
\def\ris{\right]}
\def\lec{\left\{}
\def\ric{\right\}}

\def\.{\mbox{ \tiny{$^\bullet$} }}

\def\eps{\varepsilon}

\def\epso{\eps_{\scriptscriptstyle 0}}

\def\muo{\mu_{\scriptscriptstyle 0}}

\def\ko{k_{\scriptscriptstyle 0}}

\def\ux{\hat{\#u}_{\,x}}
\def\uy{\hat{\#u}_{\,y}}
\def\uz{\hat{\#u}_{\,z}}

\def\calA{{\cal A}}
\def\calB{{\cal B}}

\begin{document}

\begin{center}

\LARGE{ {\bf  Simultaneous existence of amplified and attenuated Dyakonov surface waves}
}
\end{center}
\begin{center}
\vspace{10mm} \large

 Tom G. Mackay\footnote{E--mail: T.Mackay@ed.ac.uk.}\\
{\em School of Mathematics and
   Maxwell Institute for Mathematical Sciences\\
University of Edinburgh, Edinburgh EH9 3FD, UK}\\
and\\
 {\em NanoMM~---~Nanoengineered Metamaterials Group\\ Department of Engineering Science and Mechanics\\
Pennsylvania State University, University Park, PA 16802--6812,
USA}\\
 \vspace{3mm}
 Akhlesh  Lakhtakia\\
 {\em NanoMM~---~Nanoengineered Metamaterials Group\\ Department of Engineering Science and Mechanics\\
Pennsylvania State University, University Park, PA 16802--6812, USA}

\normalsize

\end{center}

\begin{center}
\vspace{15mm} {\bf Abstract}

\end{center}

The propagation of  Dyakonov surface waves guided by the planar interface of (i) an isotropic dielectric  material and (ii) a  homogenized uniaxial dielectric composite material comprising both passive and active component materials was theoretically
investigated, under the assumption that the optic  axis of the uniaxial partnering material lies wholly in the interface plane.
For a certain range of the volume fraction of the active component material,
the uniaxial partnering material is neither wholly dissipative nor wholly active and  the Dyakonov surface waves propagating in certain directions amplify but the Dyakonov surface waves propagating in other directions attenuate.

\section{Introduction}

Major nanotechnological developments are currently spurring the conceptualization and
occasional realization of metamaterials, which are
engineered materials with exotic optical properties  \c{Walser}. Notably, composite materials
comprising both active and dissipative component materials seem to transcend merely the ability to
 overcome dissipation \c{Sun_APL,Wuestner,Dong_APL,Strangi}. As an example, a  random mixture of  aligned
 spheroidal particles, both active and dissipative, may be theoretically equivalent
to a uniaxial dielectric material  in which a plane wave  propagating in a particular direction
can   exhibit amplification but a second plane wave propagating in a different direction gets attenuated \c{ML_PRA}. 
Furthermore,  a multilaminate structure comprising alternate  active and dissipative component layers  may be theoretically
equivalent to a birefringent material in which planewave polarization states can be controlled \c{Fan_PRL}.
Homogenization formalisms also predict the possibilities of (i) an isotropic chiral material in which  circularly polarized light
of one handedness is amplified but    circularly polarized light of the other
handedness is attenuated \c{ML_chiral}, and (ii)
a biaxial dielectric material which amplifies incident light of one linear polarization state
but absorbs incident light of the orthogonal polarization state
\c{ML_JO_2017}.

In this short paper, we focus on
Dyakonov surface waves  \c{Marchevskii,Dyakonov88,Walker98} 
guided by the planar interface of an isotropic dielectric material and an anisotropic dielectric composite material
comprising both active and dissipative component materials. Previous theoretical as well
as experimental investigations have shown that, if both partnering materials are nondissipative (and inactive),
 Dyakonov-surface-wave propagation  is generally possible    only for  propagation directions in very small angular sectors
 \c{Crasovan08,Takayama_exp}. Nevertheless, Dyakonov surface waves are attractive as they   offer considerable 
 potential for long-range optical communications \cite{Takayama2012,Noginov}.  Furthermore, the range of propagation directions may be greatly  increased  if the anisotropic partnering material is  either a hyperbolic material \cite{Vuk,Zhang} or a periodically nonhomogeneous material \cite{LP2007}, or if at least one of the two partnering materials is   dissipative  \c{ML_IEEE_Photonics}.

We report here on the prospects for amplification and attenuation of Dyakonov surface waves, by
 solving the corresponding canonical boundary-value problem \c{ESW_book} in which an inactive isotropic  material occupies the half-space $z<0$ while 
 an anisotropic  engineered material, comprising both active and dissipative component materials, occupies the half-space $z>0$.
The unit vectors parallel to the Cartesian axes are written as $ \ux$, $\uy$, and $\uz$. The free-space wave number is written as $\ko = \omega \sqrt{\epso \muo}$
with $\omega$ being the angular frequency and $\epso$ and $\muo$ being the permittivity and permeability of free space, respectively.

\section{Theoretical Preliminaries}

Both Marchevski\u{i} \textit{et al.} \c{Marchevskii} and Dyakonov \cite{Dyakonov88} have provided the    formalism   for  surface-wave propagation guided by the planar  interface of  an isotropic dielectric material and
a uniaxial dielectric material whose optic axis  is  oriented to lie wholly in the interface plane. We are content to provide
only essential details here, while referring the 
 interested reader
to more recent literature for greater illumination \cite{Furs2005, Crasovan08,ESW_book}.

The uniaxial  partnering material, labeled $\mathcal{A}$, is taken to occupy 
 the half-space $z>0$. Its relative permittivity dyadic is expressed as
\begin{equation}
\=\eps_\mathcal{A} = \eps_\mathcal{A}^{\rm s} \=I + \le
\eps_\mathcal{A}^{\rm t} - \eps_\mathcal{A}^{\rm s} \ri \,
\hat{\#{u}}_{\rm } \, \hat{\#{u}}_{\rm }\,, \l{Ch4_eps_uniaxial}
\end{equation}
with $\=I$ being the identity dyadic \cite{Chen} and
the optic axis being parallel to the
unit vector
\begin{equation}
\hat{\#{u}}_{\rm } = \ux \cos\psi+    \uy\sin\psi
\end{equation}
lying wholly in the $xy$ plane.
The eigenvalue $\eps_\mathcal{A}^{\rm s}$ governs the propagation of \textit{ordinary}
plane waves, while the  eigenvalues
 $\eps_\mathcal{A}^{\rm t}$
and  $\eps_\mathcal{A}^{\rm s}$  jointly govern the propagation of \textit{extraordinary}
plane waves, in material $\mathcal{A}$ \cite{Chen,BW}.
 The half-space $z<0$ is occupied by an isotropic  dielectric material, labeled $\mathcal{B}$, 
 whose relative permittivity 
 is denoted by
$\eps_\mathcal{B}$.

Without loss of generality, the Dyakonov   surface wave
is considered to propagate parallel to $\ux$ in the $xy$ plane with $\ko{q}$ denoting the surface wavenumber.
Thus, the direction of propagation is oriented at angle $-\psi$ with respect to the optic axis.
The normalized propagation constant $q$,
${\rm Re}\lec q \ric>0$, is determined by solving the dispersion equation \cite{Dyakonov88}
\begin{eqnarray} \l{DE}
&&
  \eps^s_\calA  \le \eps_\calB\, \alpha_{\calA 1}+ \eps^s_\calA\, \alpha_{\calB} \ri
 \le  \alpha_{\calB} + \alpha_{\calA 2} \ri  \tan^2 \psi
\nonumber \\
&&\qquad =\alpha_{\calA 1} \le  \alpha_{\calB} +\alpha_{\calA 1} \ri
\le \eps^s_\calA \alpha_{\calB} 
\alpha_{\calA 2} +\eps_\calB\, \alpha_{\calA 1}^2 \ri\,,
\end{eqnarray}
where 
 \begin{equation} \l{a_decay_const}
\left.
\begin{array}{l}
\alpha_{\calA 1} = \sqrt{q^2 -\eps_\calA^s} \vspace{8pt}\\
\alpha_{\calA 2} = \displaystyle{
\sqrt{ \eps^t_\calA \les
q^2 \le \frac{ \cos^2 \psi}{\eps^s_\calA} +
\frac{\sin^2 \psi}{\eps^t_\calA} \ri -1 \ris}}
\end{array}
\right\}\,
\end{equation}
and 
  \begin{equation} 
  \l{b_decay_const}
\alpha_\calB = \sqrt{q^2 - \eps_\calB }
\end{equation}
are such that
${\rm Re}\lec\alpha_{\calA 1}\ric>0$, ${\rm Re}\lec\alpha_{\calA 2}\ric>0$, and ${\rm Re}\lec\alpha_\calB\ric>0$. 
The Dyakonov surface wave grows in magnitude as it propagates  if ${\rm Im}\lec q \ric < 0$ but diminishes
if  ${\rm Im}\lec q \ric > 0$.

According to Eq.~\r{DE}, if a Dyakonov surface wave can propagate for $\psi = \psi^\star$, then
Dyakonov-surface-wave propagation is also possible for $\psi = - \psi^\star$ and $\psi = \pi \pm \psi^\star$. Parenthetically, these Dyakonov-surface-wave symmetries do not hold if the isotropic dielectric material occupying the 
half-space
$z<0$  is endowed with a surface admittance \c{ML_JOSAB_TI}.

\section{Numerical studies}

\subsection{Partnering materials}

 In order to  investigate the combined effects of dissipation and amplification
 in the partnering materials on Dyakonov-surface-wave propagation, 
  material
 $\mathcal{A}$ was taken to be an engineered composite material
 comprising isotropic dielectric component materials 
$\mathcal{A}a$ and $\mathcal{A}b$. Their relative permittivities are denoted
by $\eps_{\mathcal{A}a}$ and $\eps_{\mathcal{A}b}$ and their volume fractions by $f_{\mathcal{A}a}$ and $f_{\mathcal{A}b}=1-f_{\mathcal{A}a}$, respectively. The degree of anisotropy may be maximized by assuming  that the component materials are randomly distributed as  aciculate particles   oriented parallel to  $\hat{\#{u}}$ \cite{WLM1993}. The Bruggeman homogenization formalism then provides the following estimates for the relative permittivity parameters of  material
 $\mathcal{A}$  \c{Giant_anisotropy}:
\begin{equation} \l{eps_na}
\left.
\begin{array}{l}
 \eps_\mathcal{A}^{\rm s}  = \displaystyle{\frac{1}{2} \Big[ \le f_{\mathcal{A}b}-f_{\mathcal{A}a} \ri \le \eps_{\mathcal{A}b} - \eps_{\mathcal{A}a} \ri }\\
 \hspace{10mm}
  \displaystyle{ \left.
 + \sqrt{\les \le f_{\mathcal{A}b} -f_{\mathcal{A}a} \ri \le \eps_{\mathcal{A}b} - \eps_{\mathcal{A}a} \ri \ris^2 + 4 \eps_{\mathcal{A}a} \eps_{\mathcal{A}b}} \, \ris} \vspace{8pt} \\
\eps_\mathcal{A}^{\rm t} = f_{\mathcal{A}a} \eps_{\mathcal{A}a} + f_{\mathcal{A}b} \eps_{\mathcal{A}b}
\end{array}
\right\}\,.
\end{equation}
In a similar vein, a laminated composite material could yield the desired high degree of anisotropy for  material
 $\mathcal{A}$  \cite{BW}.

For component material $\mathcal{A}a$ we chose  a dissipative   material with relative permittivity  $\eps_{\mathcal{A}a} = 3 + 0.05i$.
For component material $\mathcal{A}b$ we chose  an active   material with relative permittivity  $\eps_{\mathcal{A}b} =  2 - 0.03 i $.
 This value of $\eps_{\mathcal{A}b}$ sits within the range typically employed for active components of metamaterials in the visible regime. As a  specific illustration,
a mixture of  Rhodamine 800  and  Rhodamine 6G  exhibits a relative permittivity with imaginary part in the range
$\le -0.15, -0.02 \ri$ and real part in the range $\le 1.8, 2.3 \ri$ across the frequency range 440--500 THz,
depending upon the relative concentrations  and the external pumping rate \c{Sun_APL}.

\begin{figure}[!htb]
\centering
\includegraphics[width=7.6cm]{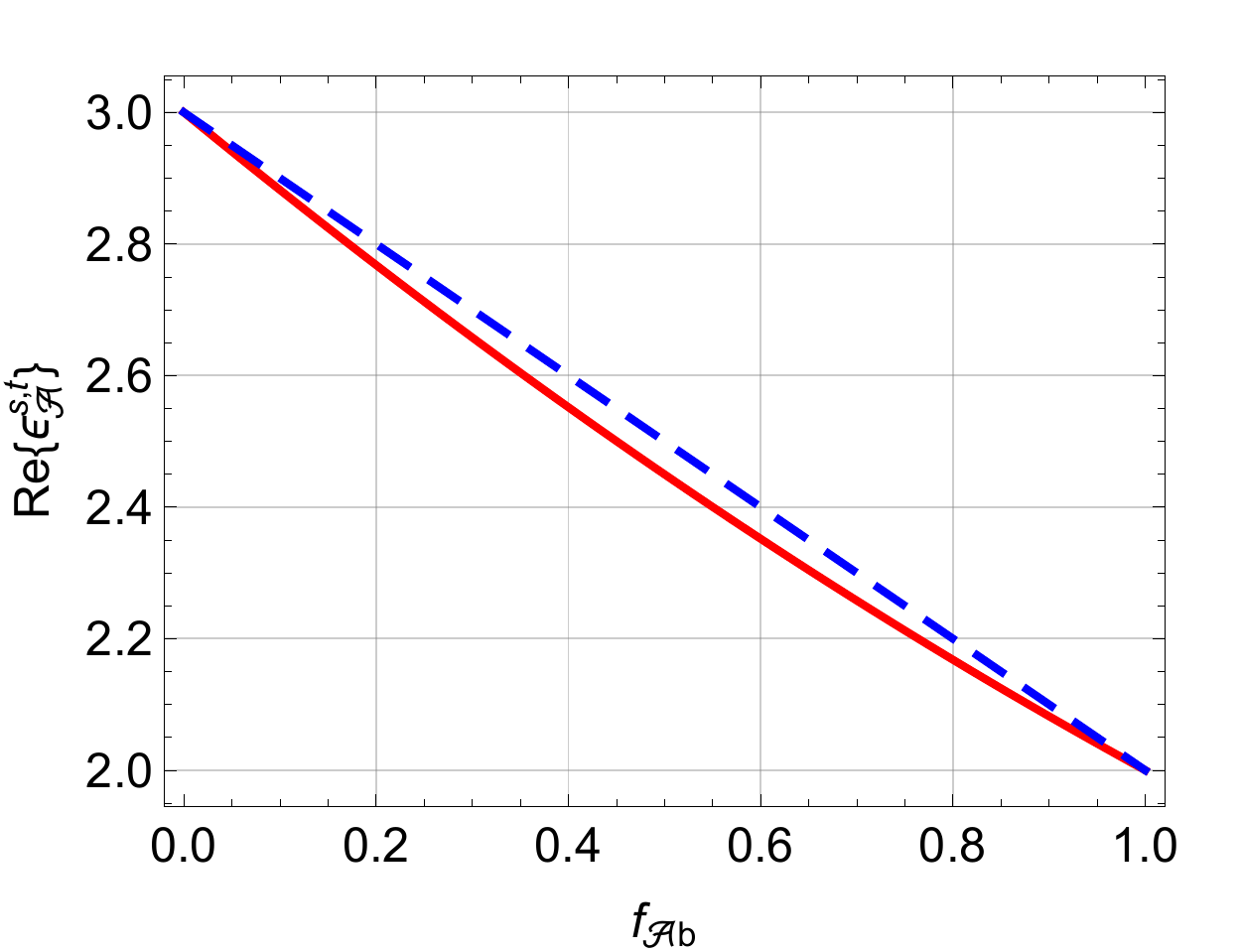}  \hfill
\includegraphics[width=7.6cm]{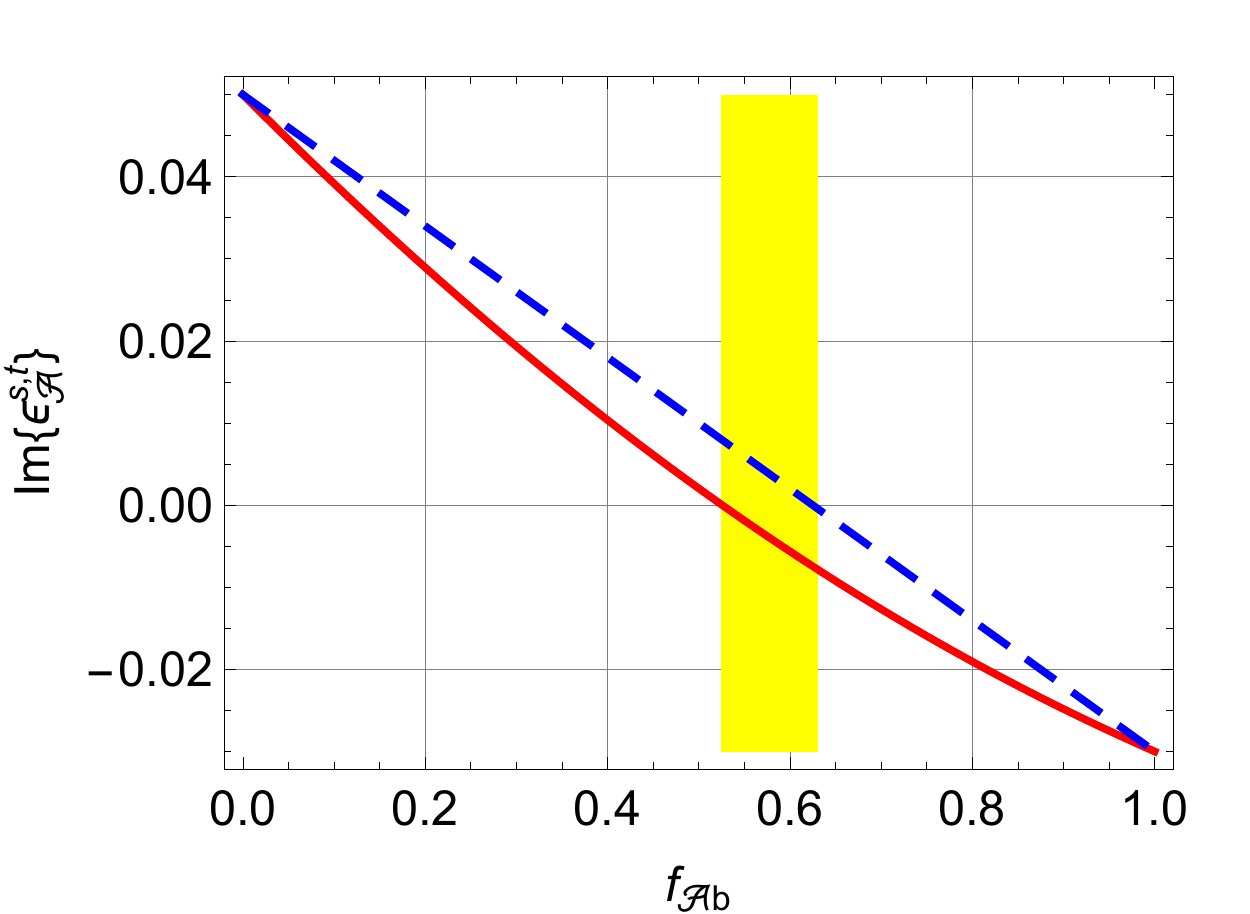}
  \caption{\label{Figure1} 
Real and imaginary parts of  $\eps_\mathcal{A}^{\rm s}$ (red solid curves) and $\eps_\mathcal{A}^{\rm t}$ (blue dashed curves) plotted against volume fraction $f_{\mathcal{A}b}$. The range $f_{\mathcal{A}b} \in \le 0.52, 0.63  \ri$ for which $\mbox{Im} \lec \eps_\mathcal{A}^{\rm s} \ric$ 
and $\mbox{Im} \lec \eps_\mathcal{A}^{\rm t} \ric$ have opposite signs is indicated in yellow.
 }
\end{figure}

The Bruggeman estimates of   $\eps_\mathcal{A}^{\rm s}$ and $\eps_\mathcal{A}^{\rm t}$
are plotted versus volume fraction $f_{\mathcal{A}b}$  in Fig.~\ref{Figure1}. Both $\mbox{Im} \lec  \eps_\mathcal{A}^{\rm s} \ric>0$ and $\mbox{Im} \lec  \eps_\mathcal{A}^{\rm t} \ric>0$    for $f_{\mathcal{A}b} < 0.52$, so that the composite material $\mathcal{A}$  is    dissipative. In contrast,  material $\mathcal{A}$ is   active   for
$f_{\mathcal{A}b} > 0.63$ since  both  $\mbox{Im} \lec  \eps_\mathcal{A}^{\rm s} \ric<0$ and $\mbox{Im} \lec  \eps_\mathcal{A}^{\rm t} \ric<0$ then.  Most interestingly, 
 for $f_{\mathcal{A}b} \in \le 0.52, 0.63  \ri$ material $\mathcal{A}$ is neither wholly active nor wholly dissipative since
$\mbox{Im} \lec  \eps_\mathcal{A}^{\rm s} \ric < 0 $ but $\mbox{Im} \lec  \eps_\mathcal{A}^{\rm t} \ric > 0 $
then.
In the next section, Dyakonov-surface-wave propagation is investigated for   $f_{\calA b}=0.4$, $0.58$, and $0.8$,
 material $\calA$ then being wholly dissipative, neither wholly active nor wholly dissipative, and wholly active, respectively.

The following values of the relative permittivity of material $\calB$ were chosen for numerical investigation: $\eps_\mathcal{B} = 2.5$, $2.5 + 0.01i$, and $2.5 + 0.2i$.

\subsection{Surface-wave analysis}

\begin{figure}[!htb]
\begin{center}\includegraphics[width=7.6cm]{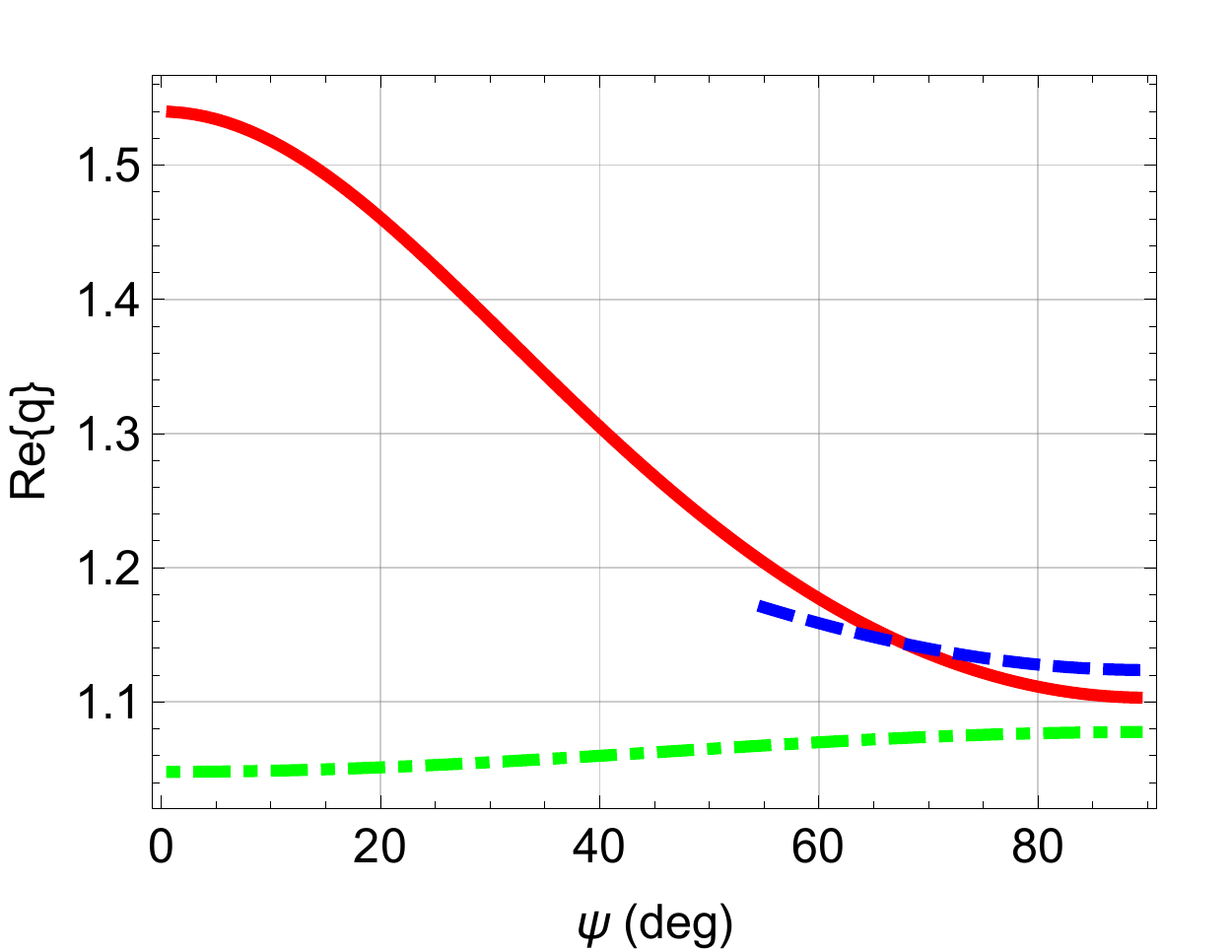} \hfill
\includegraphics[width=7.6cm]{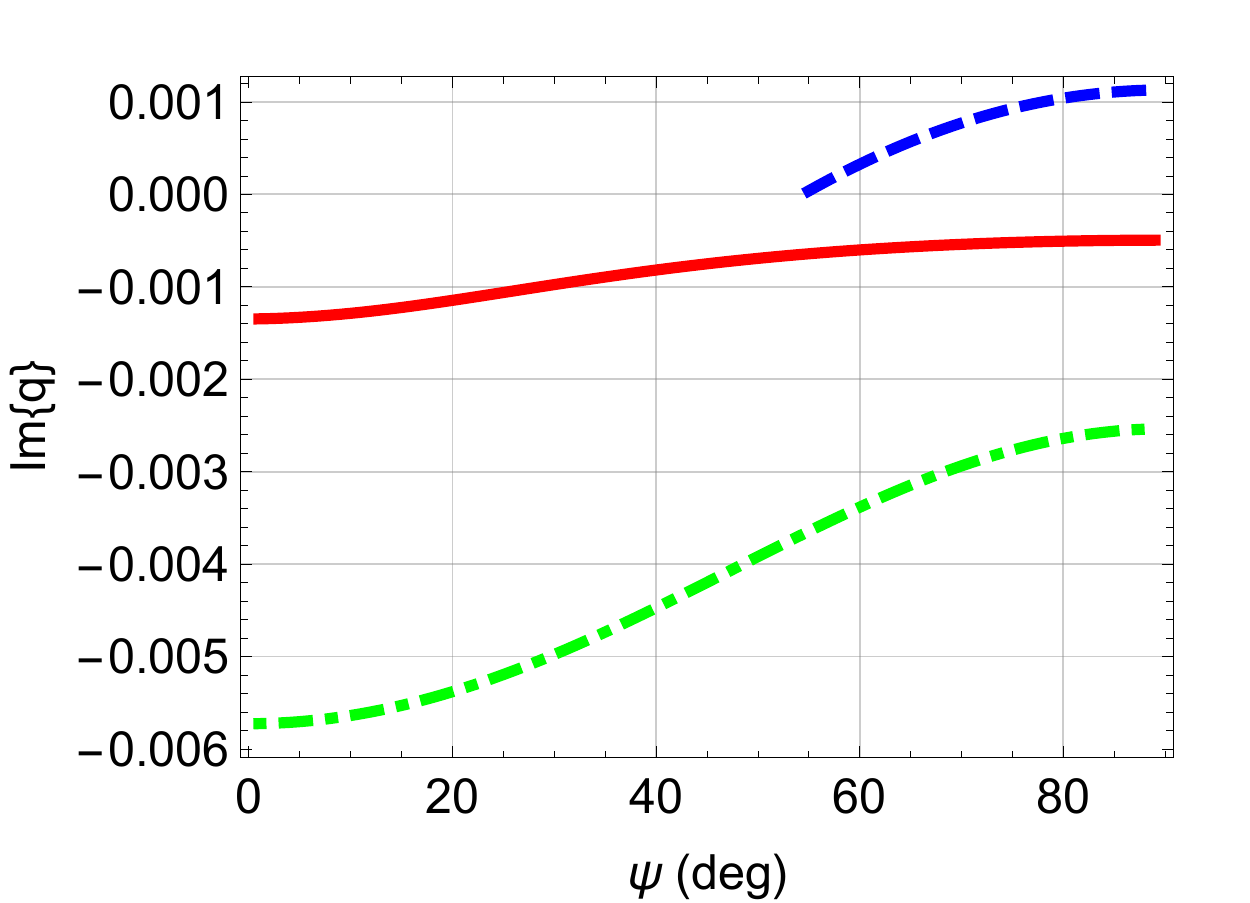} \vspace{0mm}  
\end{center}
 \caption{${\rm Re}\lec{q}\ric$ and ${\rm Im}\lec{q}\ric$    plotted against   $\psi\in[0^\circ,90^\circ]$ for $f_{\calA b} = 0.4 $  (blue dashed curves), $0.58$ (red solid curves), and $0.8$ (green broken dashed curves), when $\eps_\mathcal{B} =  2.5 $.
 } \label{Figure2}
\end{figure}

Let us begin with the case in which material $\calB$ is   nondissipative: $\eps_\mathcal{B} = 2.5$.
 The real and imaginary parts of the normalized propagation constant $q$, as delivered from Eq.~\r{DE}, are plotted against the  angle $\psi$  for $f_{\calA b} \in \lec 0.4, 0.58, 0.8 \ric$ in Fig.~\ref{Figure2}. 
 Dyakonov-surface-wave propagation is possible for $\psi \in \le 55^\circ, 90^\circ \ri$ when $f_{\calA 2} = 0.4$, but for
$\psi \in \le 0^\circ, 90^\circ\ri$ when
$f_{\calA b} \in\lec0.58,0.8\ric$.
From the plots of $\mbox{Im} \lec q \ric$,  we infer that the Dyakonov surface wave 
is attenuated when $f_{\calA b} = 0.4 $
but
amplified when
$f_{\calA b} \in\lec0.58,0.8\ric$, regardless of direction of propagation (as given by $\psi$).

Parenthetically, the  large angular existence domains (i.e., ranges of values $\psi$ for which Dyakonov-surface-wave propagation is possible)  in Fig.~\ref{Figure2} are attributable to the dissipative and active nature of material $\calA$. Indeed,  previous studies have revealed that if the partnering materials are characterized by relative permittivity parameters with  nonzero imaginary parts then large 
angular existence domains can be obtained for Dyakonov surface waves \c{ML_IEEE_Photonics}. In contrast, when both
partnering materials are taken to be nondissipative and nonactive, the angular existence domains are minuscule \c{Crasovan08,Takayama_exp}.

\begin{figure}[!htb]
\begin{center}\includegraphics[width=7.6cm]{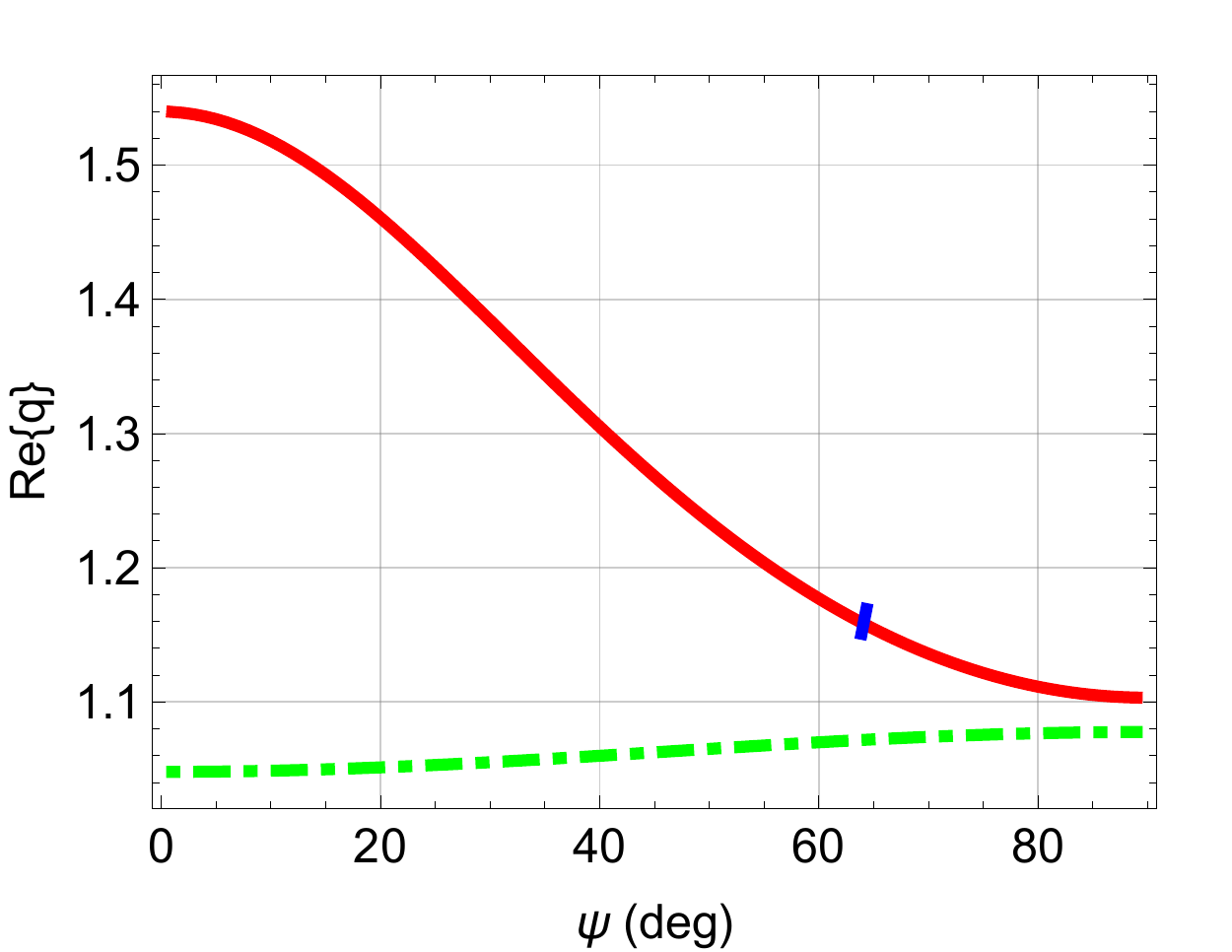} \hfill
\includegraphics[width=7.6cm]{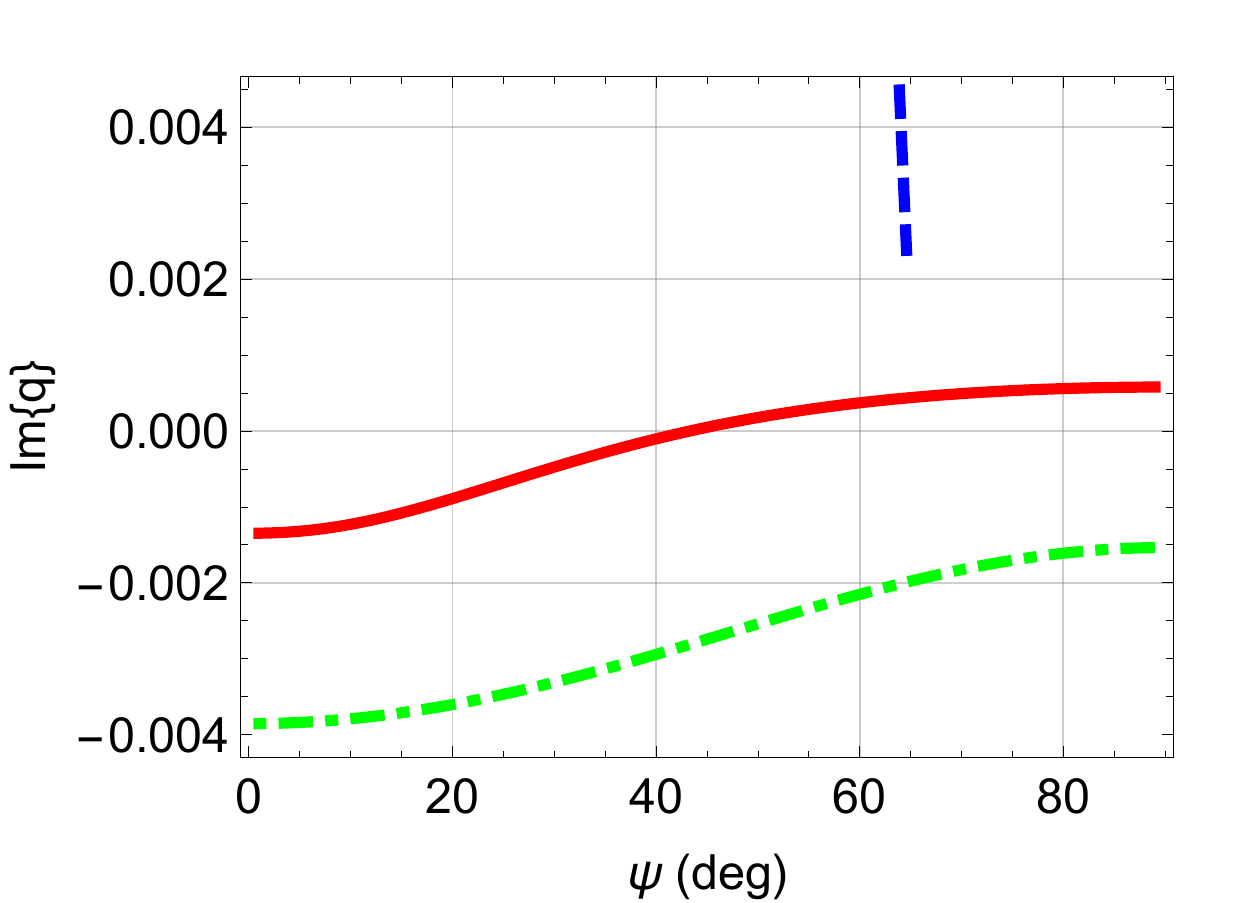} \vspace{4mm}  
\end{center}
 \caption{As Fig.~\ref{Figure2} except that $\eps_\mathcal{B} =  2.5 + 0.01 i $.
 } \label{Figure3}
\end{figure}

Next we turn to the case in which material $\calB$ is  weakly dissipative: $\eps_\mathcal{B} = 2.5 + 0.01i$. Plots of the 
  real and imaginary parts of  $q$ 
  versus $\psi$ are provided in Fig.~\ref{Figure3}
    for $f_{\calA b} \in \lec 0.4, 0.58, 0.8 \ric$.
The Dyakonov  surface wave exists only for $\psi \in \le 63.9^\circ, 64.6^\circ \ri$ when  $f_{\calA b} = 0.4 $, but for
$\psi \in \le 0^\circ, 90^\circ\ri$ when
$f_{\calA b} \in\lec0.58,0.8\ric$.
   The tiny angular existence domain for $f_{\calA b} = 0.4 $ is typical for Dyakonov surface waves when both partnering materials are nondissipative and inactive \c{Crasovan08,Takayama_exp}.
From the plots of $\mbox{Im} \lec q \ric$ we infer that the Dyakonov surface wave 
is attenuated when $f_{\calA b} = 0.4 $
and
amplified when
$f_{\calA b} = 0.8$, regardless of the direction of propagation. But the 
case $f_{\calA b} = 0.58 $ combines both  characteristics since  $\mbox{Im} \lec q \ric < 0$ for $\psi \in \le 0^\circ, 43^\circ \ri$
but $\mbox{Im} \lec q \ric > 0$ for $\psi \in \le 43^\circ, 90^\circ \ri$. Thus, the Dyakonov surface wave is
attenuated  for large values of $\psi$ but amplified for small values of $\psi$.

\begin{figure}[!htb]
\begin{center}\includegraphics[width=7.6cm]{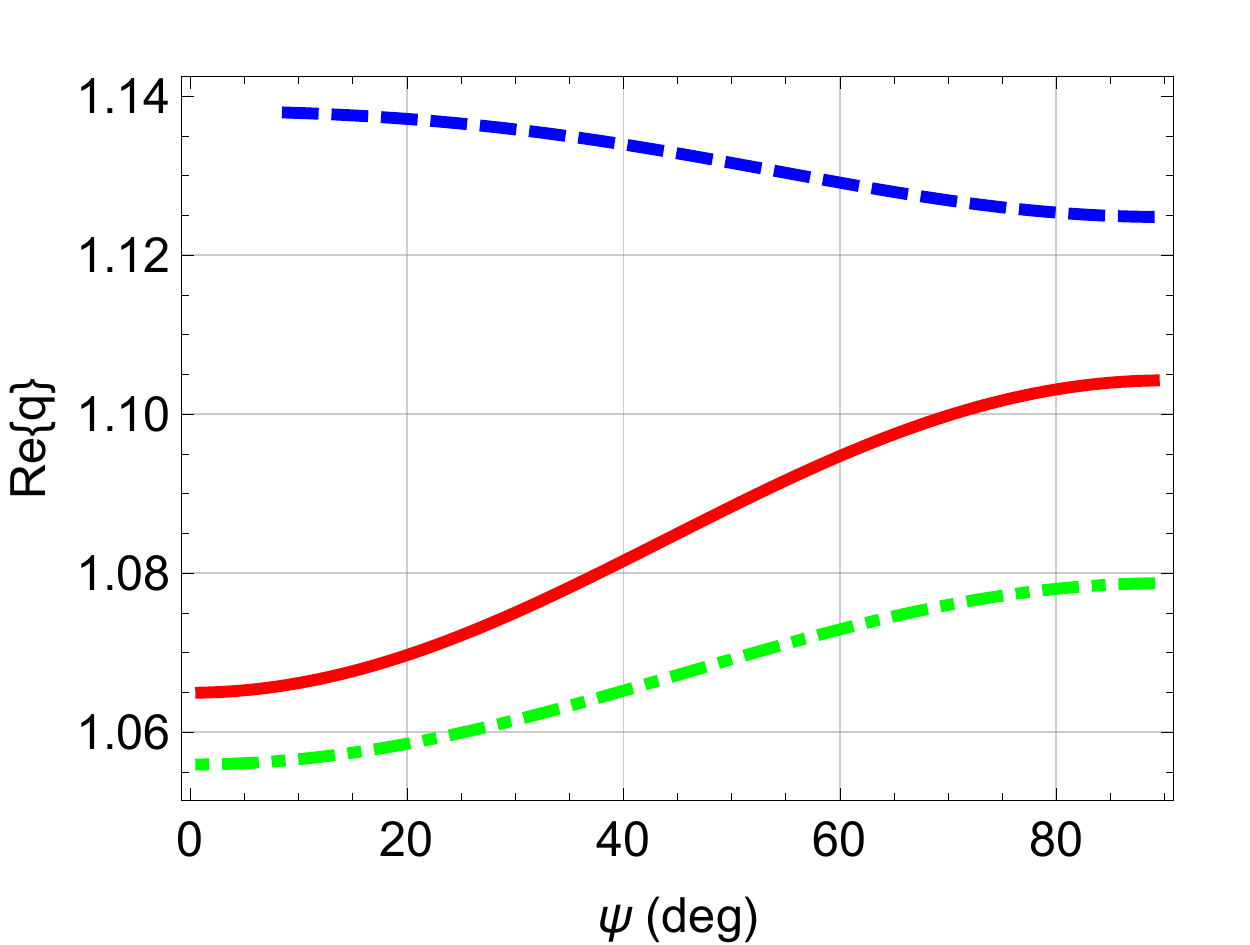} \hfill
\includegraphics[width=7.6cm]{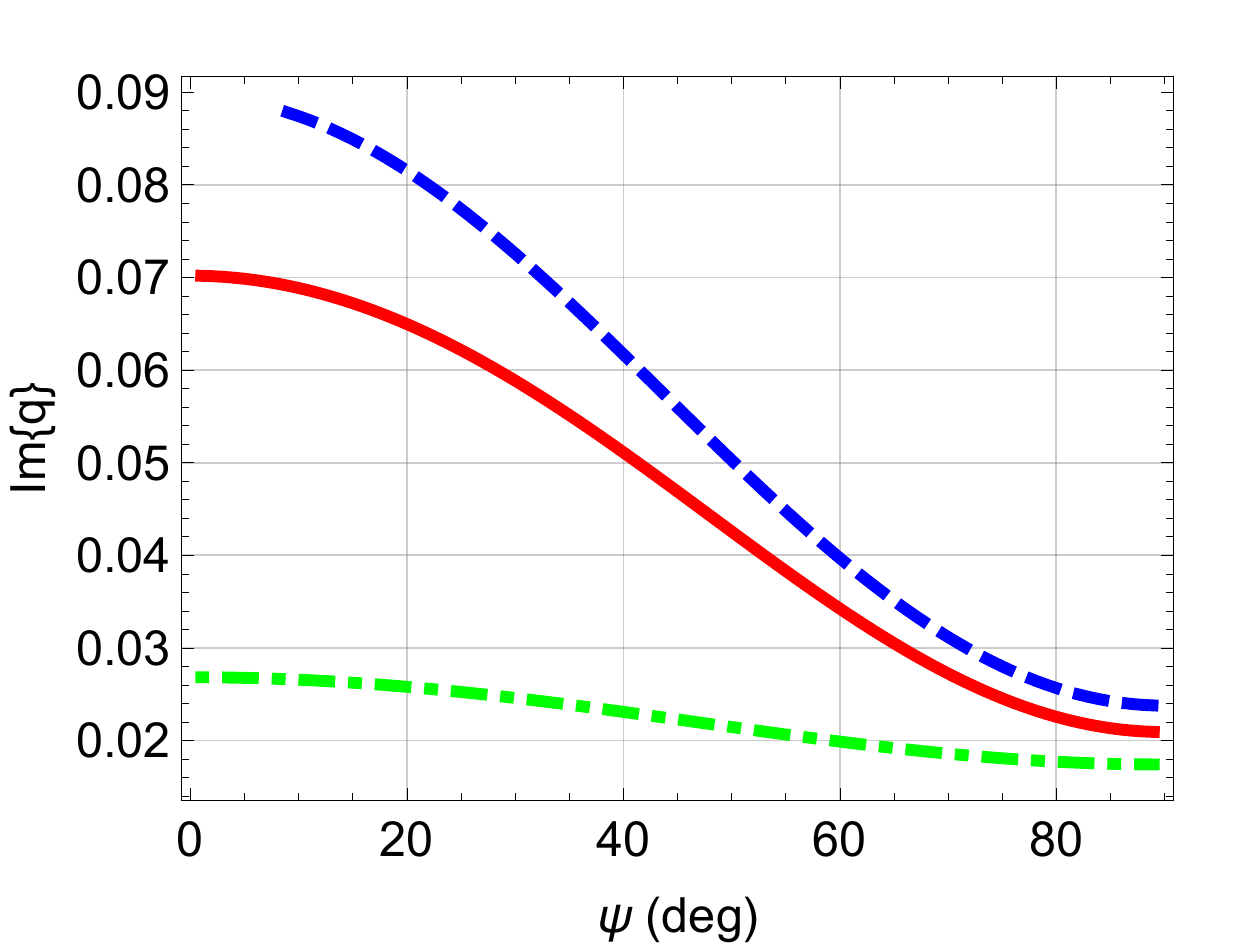} \vspace{4mm}  
\end{center}
 \caption{As Fig.~\ref{Figure2} except that $\eps_\mathcal{B} =  2.5 + 0.2 i $.
 } \label{Figure4}
\end{figure}

Lastly we consider the case of $\calB$ being  strongly dissipative: $\eps_\mathcal{B} = 2.5 + 0.2i$.  The 
  real and imaginary parts of  $q$ 
  are plotted versus $\psi$  in Fig.~\ref{Figure4}
    for $f_{\calA b} \in \lec 0.4, 0.58, 0.8 \ric$.
Dyakonov-surface-wave propagation is possible
for $\psi \in \le 9^\circ, 90^\circ \ri$ when      $f_{\calA b} = 0.4 $, but for
all directions when
$f_{\calA b} \in\lec0.58,0.8\ric$. Furthermore,  the Dyakonov surface wave 
is attenuated for all values of  $f_{\calA b}$ considered, regardless of direction of propagation.

\section{Closing remarks}

Dyakonov surface waves guided by the interface of nondissipative (and inactive) materials are generally restricted to tiny angular existence domains \c{Crasovan08}. By allowing at least one of the partnering materials to be dissipative (or active),  angular existence domains can 
be increased substantially \c{ML_IEEE_Photonics}. The
 propagation of Dyakonov surface waves guided by the planar interface of  an isotropic dielectric material and  a homogenized uniaxial dielectric composite material comprising both passive and active component materials was theoretically investigated,  with  the optic axis of the uniaxial partnering material assumed to lie wholly in the interface plane. For a certain range of the volume fraction of the active component material, the uniaxial partnering material is neither wholly 
 dissipative nor wholly active, Dyakonov-surface-wave propagation can occur in every direction in the interface plane, and the Dyakonov surface waves propagating in certain directions amplify but the Dyakonov surface waves propagating in other directions attenuate. These characteristics may be  exploited for applications involving optical sensing and/or optical communications \c{Takayama2012,Noginov}.

\vspace{2mm}

\noindent {\bf Acknowledgments.}
AL is grateful to the Charles Godfrey Binder Endowment at Penn State for ongoing support of his research.

\end{document}